# THE INFLUENCE OF THE COULOMB INTERACTION IN ELECTRON BUNCH ON THE STRUCTURAL PARAMETERS DETERMINED BY TIME-RESOLVED ELECTRON DIFFRACTION

**A.A. Ischenko[1], Yu.I. Tarasov[1] and V.L. Popov[2]**

[1] Moscow Lomonosov State University of Fine Chemical Technologies,

Moscow, Vernadskogo prosp. 86, Russia

[2] Berlin University of Technology, Sekr. C8-4, 10623 Berlin, Germany

**Abstract**

Transition to temporal resolution on the order of 1 ps or less raises a number of questions associated with estimation of the accuracy of the dynamic parameters based on the analysis of time-dependent scattering intensity. The use of ultrashort pulses leads to the necessity for increasing the total exposure time and lengthening the time of data acquisition. This can be to some extent mitigated by increasing the charge per pulse. Increasing in the number of electrons, however, is limited by the Coulomb repulsion between them, which leads on one hand, to distortion of the diffraction pattern, and the other hand to increase the duration of the pulse. The analytical technique for estimating the deformation of the diffraction pattern caused by the Coulomb repulsion of the electrons in the electron bunch with duration of less than 10 ps, and the influence of this effect on the accuracy of determination of the interatomic distances is developed.

## 1. Introduction

In 1927, Davisson and Germer [1] and independently Thomson and Reid [2] discovered the phenomenon of diffraction of electrons on crystals. This finding relates to the "static" diffraction. After the first experiments performed by Mark and Wierl in 1930 [3,4] the method of electron diffraction remained conceptually unchanged until the early 80's of the last century. Only 50 years later it became possible to introduce fourth dimension in the technique - time, introducing the concept of structural dynamics and research in 4D space - time continuum by Time-Resolved Electron Diffraction (TRED). A detailed description of the history of the development is given in the review articles [5-7] and in the books [8-10].

The nuclear motion as a chemical reaction unfolds can be probed using TRED. Breaking of chemical bonds, their formation and change in the geometry of the molecule occurs at a rate of ~1000 m/s (see, e.g., [11]). Consequently, for the registration of nuclear dynamics the time resolution of <100 fs is required. Regardless of whether the molecule is isolated or not the ultrafast changes involve coordinated rearrangement of the electron and nuclear subsystems of reacting molecules.

According to the transition state theory [12-15], for the unimolecular reaction we have a frequency factor equal to ($kT / h$) - namely with this frequency the transition to the final products occurs through the energy barrier of the chemical reaction [16-18]. At room temperature, its value is ~ $6 \times 10^{12}$ s$^{-1}$, which corresponds to the time of ~ 150 fs. Typical intramolecular vibrations occur in the time interval of the order of hundreds of femtoseconds. That is why, to some extent the term "ultrafast electron diffraction" is justified. In this case, it is clear that the diffraction process of electrons is determined by the accelerating voltage and single scattering event for 100-keV-electron occurs in a few as.

In this article we present the analytical technique for estimating the deformation of the diffraction pattern caused by the Coulomb repulsion of the electrons in the electron bunch with duration of less than 10 ps, and the influence of this effect on the accuracy of determination of the interatomic distances.

## 2. Theory

Transition to temporal resolution on the order of 1 ps or less raises a number of questions associated with estimation of the accuracy of the dynamic parameters based on the analysis of time-dependent scattering intensity [5-7, 19-22]. The use of ultrashort pulses leads to the necessity for increasing the total exposure time and lengthening the time of data acquisition. This can be to some extent mitigated by increasing the charge per pulse. Increasing in the number of electrons, however, is limited by the Coulomb repulsion between them, which leads on one hand, to distortion of the diffraction pattern, and the other hand to increase the duration of the pulse.

Determination of the scattering coordinates is shown in Figure 1. Analysis of distortion of diffraction pattern due to Coulomb repulsion was first performed in [23]. The authors of the paper considered the situation where the spatial pulse length, defined as $L = v\tau_e$ ($v$ – the velocity of the electrons, $\tau_e$ – pulse duration) is considered greater than the distance from the scattering region to the screen. The domain of applicability of the results [23] can be expressed by the inequality: $R_D \ll L = v\tau_e$, where $R_D$ – is a characteristic radius of the diffraction pattern in the detection plane. For the commonly used electron energies ~ 30-100 keV, this condition can be written in the form of restrictions on the duration of the electron pulse: $\tau_e \gg 10^{-10}$ s. It is clear that this condition can not be considered appropriate in the study of the dynamics of nuclei by Time-Resolved Electron Diffraction (TRED), because the relevant processes occur on a time scale of picoseconds to femtoseconds.

## 2.1. Description of the model

General scheme of the TRED experiment and the determination of the coordinates of the scattered electron are presented on Figure 1. The following assumptions were made:

(A) Since only a small fraction of electrons in the bunch get scattered [8-11, 23-25], we considered that scattered electrons only interacted with the electrons of the original (unscattered) electron bunch. Moreover, the number of the electrons in the bunch was held constant.

(B) The relative motion of the electrons in the electron bunch is determined by charge density function and the Coulomb interaction. Denoting the initial velocity of the electrons in the beam through $v_z$ for electron energy of 50 keV their speed is about 1/3 of the speed of light. Non-relativistic dynamics gives an accuracy of 10%. In our calculations, we neglect the relativistic effects and use the Galilean transformation for the transition to a reference frame which is moving together with the electron bunch.

(C) Calculations were carried out in the assumptions of small-angle scattering, which is a common approach in TRED [8-11, 24-26].

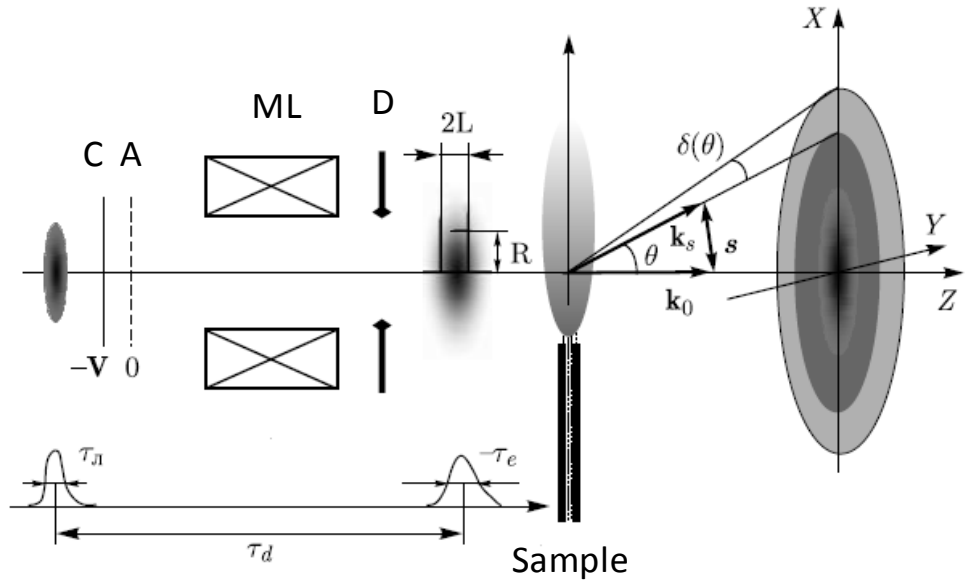

**Fig. 1.** Determination of the coordinates of the scattered electron. $\theta$ - the scattering angle, $\delta(\theta)$ – a correction to the scattering angle caused by Coulomb repulsion in the electron bunch; $k_o$ and $k_s$ - wave vectors of the incident and scattered electrons, respectively; s - momentum transfer vector in the laboratory frame XYZ; $\lambda$ - wavelength of electrons. C- cathode, A - anode, ML - Magnetic Lenses, D - Diaphragm; $\tau_L$ - the duration of the laser pulse, $\tau_e$ - electron pulse duration; $l$ - axis of the electron bunch in the direction of its motion, $R$ - axis of the electron bunch in the transverse direction; $\tau_d$ - the time delay between the excitation laser pulse and diagnosing electronic pulse.

## 2.2. Method for calculating corrections to the electron scattering angle based on the conservation of energy law

The influence of Coulomb interaction on the deflection angle of the scattered electron can be calculated using perturbation theory. According to the described model, for perform calculations it is convenient to use a reference frame moving with the electron bunch with speed $v_z$. The trajectory of the electron scattered at angle $\theta$ (see Figure 1) in the zero approximation, is a straight line extending from the point of scattering (Figure 2), along which the electron moves with a constant radial (relative to the trajectory of his original motion) speed $v_x^{(0)} \equiv v_0 = v_z \theta$. Coulomb interaction of an electron with electron of the pulse leads to a change of the velocity vector components $v_x^{(1)}, v_y^{(1)}, v_z^{(1)}$ in such a way that the components so the velocity vector components after scattering and the "Coulomb acceleration" are given by $\left(v_0 + v_x^{(1)}, v_y^{(1)}, v_z^{(1)}\right)$. Since the electron under the influence of the Coulomb repulsion from the electron bunch, most significantly accelerated at distances on the order of the transverse dimension of the bunch (much less than the distance of its maximum take-off), with the accepted accuracy, we can assume that it acquires an additional transverse velocity $v^{(1)}$ almost instantly and subsequently moves with a constant speed at an angle $[\theta + \delta^{(1)}]$, while not interacting with the original electron bunch. Velocity $v^{(1)}$ can be determined from the conservation of energy.

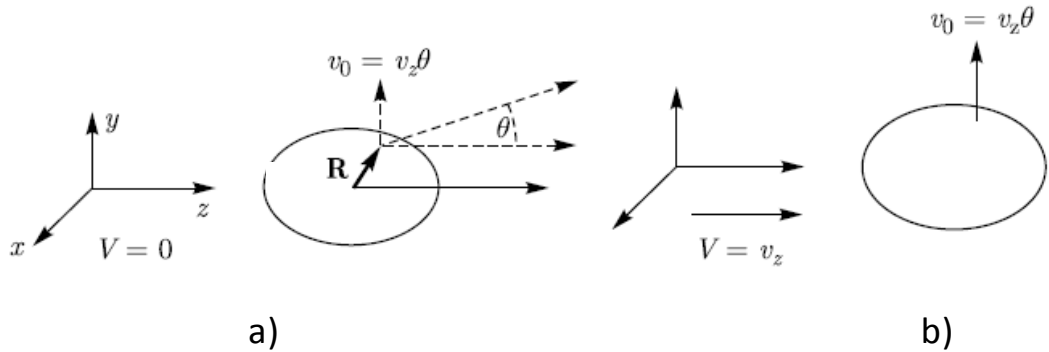

a)                                              b)

**Fig. 2.** Movement of an electron scattered in angle $\theta$ in two frames of reference: (a) – laboratory, (b) – moving together with the electron bunch: $v_0$ - initial transverse velocity of the scattered electron; $v_z$ – speed of the electron bunch.

We denote the radius vector of the electron (before the act of scattering, within the electron bunch) as **R** (Figure 2). We denote the potential energy of the scattered electron in the point defined as the distance in the plane perpendicular to the axis of motion of electron bunch as $U(\mathbf{R})$ (Figure 2). In general, all three components of the force acting from the distributed charge on the electron have the same order of magnitude. Accordingly, the projection of velocity vectors: $v_x^{(1)}, v_y^{(1)}, v_z^{(1)}$ will also be of the same order of magnitude. Energy conservation law for the scattered electron can be written as:

$$U(R) + \frac{mv_0^2}{2} = \frac{m(v_0 + v_x^{(1)})^2}{2} + \frac{m(v_y^{(1)})^2}{2} + \frac{m(v_z^{(1)})^2}{2}. \tag{1}$$

Considering, that $|v_0| \gg |v_x^{(1)}|, |v_y^{(1)}|, |v_z^{(1)}|$, in equation (1) we can neglect the terms of the second order of $v_i^{(1)}$ ($i = x, y, z$):

$$U(R) + \frac{mv_0^2}{2} \approx \frac{m(v_0 + v_x^{(1)})^2}{2} \approx \frac{mv_0^2}{2} + mv_0 v_x^{(1)}, \tag{2}$$

thus, we obtain: $U(R) = mv_0 v_x^{(1)}$. From where considering $v_0 = v_z \theta$ we get:

$$v_x^{(1)} = \frac{1}{mv_z \theta} U(R) \tag{3}$$

We note that the component of the velocity $v_y^{(1)}$ up to terms of the order of $R/R_D$ ($R$ and $R_D$ are characteristic transverse dimension of the electron bunch and the diffraction pattern, respectively) does not lead to displacement or dilution of the diffraction pattern, since it corresponds to the rotation around the axis of the diffraction pattern (see Figure 1). Component $v_z^{(1)}$ in small-angle scattering approximation, leads only to a small change in the longitudinal velocity of the electron, without changing the angle of scattering. Distortion of the diffraction pattern is only caused by radial velocity (Equation 3). Then, the correction to the scattering angle $\delta^{(1)} = v_x^{(1)} / v_z$ is equal to:

$$\delta^{(1)} = \frac{1}{\theta m v_z^2} U(R) = \frac{1}{2\theta} \frac{U(R)}{\varepsilon}, \tag{4}$$

where $\varepsilon$ is the initial kinetic energy of the electron. Formula (4) is very simple and reduces the calculation of the Coulomb distortion of the diffraction pattern to the calculation of the distribution of the electrostatic potential in the electron bunch. Interestingly, that the correction to the scattering angle in the equation (4) is inversely proportional to the scattering angle itself, i.e. make such distortions in the diffraction pattern, which can not be eliminated by scaling interatomic distances.

Obviously, the electrons scattered in different parts of the bunch, have, in general, different potential energies. According to equation (4), they undergo different deviation. In the next section we will show that the distortion of the diffraction pattern is determined in the first-order perturbation theory, only by the average value of the potential energy <U(R)>. The latter depends on the detailed form of electron density distribution. The most realistic is a Gaussian distribution:

$$\rho(R) = \frac{q}{(2\pi)^{3/2} R^2 l} \exp\left(-\frac{g^2}{2}\right) \tag{5}$$

where

$$g = \left(\frac{x^2}{R^2} + \frac{y^2}{R^2} + \frac{z^2}{l^2}\right)^{1/2}, \tag{6}$$

$q$ is the total charge of the electron bunch, $R$ and $l$ – half-width of distribution in the transverse and longitudinal directions of the electron bunch, and $x, y, z$ - components of vector $R$. The potential of such a distribution at $R = 0$ can be calculated analytically [27]. In SI units, it has the form:

$$\varphi(0) = \sqrt{\frac{2}{\pi}} \frac{k_c q}{\sqrt{l^2 - R^2}} \operatorname{arcosh} \frac{l}{R}, \quad for\ l > R \tag{7}$$

$$\varphi(0) = \sqrt{\frac{2}{\pi}} \frac{k_c q}{\sqrt{R^2 - l^2}} \arccos \frac{l}{R}, \quad for\ l < R \tag{8}$$

where $k_C = 1/4\pi\varepsilon_0 \approx 9 \cdot 10^9 (Nm^2/C^2)$ – Coulomb's constant.

For a spherically symmetric distribution ($l = R$) we get:

$$\varphi(0) = \sqrt{\frac{2}{\pi}} \frac{k_c q}{R}. \tag{9}$$

It is useful to keep in mind the following asymptotic formula for the case of a strongly elongated distribution of electron density ($l >> R$):

$$\varphi(0) = \sqrt{\frac{2}{\pi}} \frac{k_c q}{l} \ln \frac{2l}{R} \tag{10}$$

and strongly oblate ($l << R$) distributions of the electron density:

$$\varphi(0) = \sqrt{\frac{\pi}{2}} \frac{k_C q}{R}. \qquad (11)$$

We are interested in the average value of the potential $\varphi$, which can be estimated, for $0 < l/R < 10$, as:

$$\Phi = (0{,}70 \pm 0{,}02)\, \varphi(0). \qquad (12)$$

Formulas (4) and (7) – (11) completely determine the distortion of the diffraction pattern due to the Coulomb interaction for arbitrary electron pulse duration. Figure 3a shows the dependence of the potential at the center of an ellipsoidal Gaussian distribution of charge on the parameter of its elongation $l/R$ (at constant $q$ and $R$). The strongest repulsion of the electrons is observed in the limit of ultrashort bunches. Figure 3b shows the dependence of the potential on the radius of the electron bunch (with a constant spatial length $l$ and charge $q$). The decrease of the radius leads to a sharp increase of the potential $\varphi(0)$, and, consequently, an additional increase in the scattering angle. Figure 3 shows a relative increase in the potential at the center of an electron bunch, compared to a spherically symmetric charge distribution within the bunch.

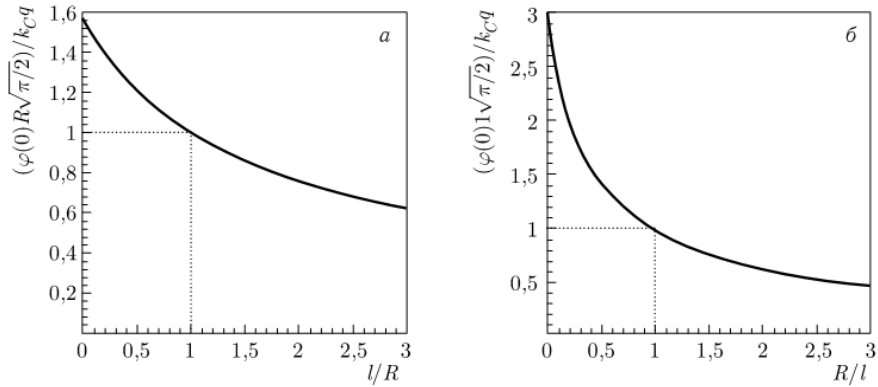

**Fig. 3.** Dependence of the potential at the center of an ellipsoidal Gaussian charge distribution on the parameter of its elongation $l/R$: (a) at constant $R$; (b) at a constant $l$. The dashed lines show the parameters for a spherically symmetric charge distribution.

**2.3. Estimation of the distortion of the measured parameters of the molecules.** We define the reduced molecular component of the electron scattering intensity as $M(s)$ (see, e.g., [25]) as $M(\theta, r)$, where $r$ is internuclear distance. $s = (4\pi/\lambda)\sin(\theta/2)$, $\lambda$ - electron wavelength. For simplicity we consider scattering by diatomic molecules.

Considering the Coulomb repulsion, electrons, which should have been scattered in angle $\theta$, will be scattered at an angle $[\theta + \delta^{(1)}(\theta)]$, which leads to a distortion of the scattering intensity:

$$\widetilde{M}(\theta,r) = M\left(\theta - \delta^{(1)}(\theta), r\right) \cdot \left(1 - \frac{\partial \delta^{(1)}(\theta)}{\partial \theta}\right) = \qquad (13)$$
$$M(\theta,r) - \frac{\partial M}{\partial \theta} \delta^{(1)}(\theta) - M(\theta,r) \frac{\partial \delta^{(1)}(\theta)}{\partial \theta} \ .$$

where $\widetilde{M}$ – is intensity of distorted diffraction pattern, or, taking into account (4):

$$\widetilde{M}(\theta,r) = M(\theta,r) + \frac{U(R)}{2\varepsilon}\left(-\frac{\partial M}{\partial \theta} \cdot \frac{1}{\theta} + \frac{M(\theta,r)}{\theta^2}\right) \ . \qquad (14)$$

This distribution depends on the point $R$, at which there was scattering. To get the final result the distribution must be averaged over $R$:

$$\left\langle \widetilde{M}(\theta,r) \right\rangle = M(\theta,r) + \frac{\langle U(R) \rangle}{2\varepsilon}\left(-\frac{\partial M}{\partial \theta} \cdot \frac{1}{\theta} + \frac{M(\theta,r)}{\theta^2}\right) \ . \qquad (15)$$

As can be seen from equation (15), the first order perturbation theory suggests that the distortion of the diffraction pattern is determined only by the average value of the potential energy of the electron $<U(R)>$.

When processing the diffraction pattern this distribution should be compared to the reference function $M(\theta, r)$ with some unknown value $\tilde{r} = r - \delta r$:

$$M(\theta, \tilde{r}) = M(\theta, r) - \frac{\partial M}{\partial r} \delta r \qquad (16)$$

Minimizing the quadratic discrepancy:

$$Q = \int_{\theta_{min}}^{\theta_{max}} \left[\left\langle \widetilde{M}(\theta,r)\right\rangle - M(\theta,\tilde{r})\right]^2 d\theta \ , \qquad (17)$$

we determine:

$$\delta r = \frac{\langle U(R) \rangle}{2\varepsilon} \cdot \frac{\int_{\theta_{min}}^{\theta_{max}} \left(-\frac{\partial M}{\partial \theta} \cdot \frac{1}{\theta} + \frac{M}{\theta^2}\right) \frac{\partial M}{\partial r} d\theta}{\int_{\theta_{min}}^{\theta_{max}} \left(\frac{\partial M}{\partial r}\right)^2 d\theta} \quad . \tag{18}$$

where $\theta_{min}$ and $\theta_{max}$ are the minimum and maximum scattering angles of the acquired diffraction pattern, respectively.

We take into account that electron scattering on the potential of the molecule occurs within ~ 1 as, that is several orders of magnitude less than the duration of electron pulse, $\tau_e$. Therefore, to estimate the integrals in (18) we can use the time-independent, "instantaneous" value of $M(\theta, r)$ at some time $t \in \tau_e$. Then, for isotropically oriented in space molecules we can use the approximation [6,7]:

$$M(\theta, r) \sim \frac{\sin(sr)}{sr} \tag{19}$$

For estimation of distortion of interatomic distances (equation 18) we assume that $s \approx \frac{2\pi}{\lambda}\theta = k\theta$. We also assume that function (19) is oscillating very fast and for determination of its derivative it is sufficient to differentiate only the numerator of the equation. With those assumptions we obtain:

$$\delta r = r \frac{\langle U(R) \rangle}{2\varepsilon} \cdot \frac{\int_{\theta_{min}}^{\theta_{max}} \left(-\cos^2 k\theta r + \sin k\theta r \cdot \cos k\theta r\right) d\theta}{\int_{\theta_{min}}^{\theta_{max}} \theta^2 \cos^2 k\theta r \, d\theta} \tag{20}$$

Since the interval of scattering angles usually fulfilles inequality $(s_{max} - s_{min})r \gg 1$ [9,10], the second term in the numerator can be neglected (the mean value is zero). Substituting in the equation 20, $\cos^2 sr$ by its mean value of ½ and integrating, we obtain:

$$\frac{\delta r}{r} = \frac{\langle U(R) \rangle}{2\varepsilon \tilde{\theta}^2} \quad , \tag{21}$$

where

$$\tilde{\theta}^2 = \theta_{max}^2 + \theta_{max}\theta_{min} + \theta_{min}^2 \quad . \tag{22}$$

Therefore, as follows from the equation 21, the relative correction becomes smaller for larger values of $\theta_{max}$ and $\theta_{min}$. Electron bunch must contain the number of electrons, which is no more than a certain value $N_{max}$, at which the uncertainty (21) does not exceed the maximal value of the experimental error. We estimate this critical value $N$ for the case of ultrashort pulses, $l < R$. For the potential of an electron within the electron bunch we should utilize (11). If we determine the accuracy of the parameters as:

$$\frac{\delta r}{r} = \kappa = \frac{\langle U(R) \rangle}{2\varepsilon \tilde{\theta}^2} \approx \frac{0,7 \cdot e\varphi(0)}{2\varepsilon \tilde{\theta}^2} = \frac{0,7 \cdot Ne^2}{2\varepsilon \tilde{\theta}^2} \sqrt{\frac{\pi}{2}} \frac{1}{R}$$

the maximum number of electrons per pulse ($N = q / e$) is determined as:

$$N_{max} = 2\sqrt{\frac{2}{\pi}} \kappa \frac{R\tilde{\theta}^2 \varepsilon}{0,70 \cdot e^2 k_C} \approx 2,3\kappa \frac{R\tilde{\theta}^2 \varepsilon}{e^2 k_C} \quad (23)$$

Substitution of typical values, $\kappa = 10^{-3}$, $\theta_{min} = 4 \cdot 10^{-2}$, $\theta_{max} = 2.5 \cdot 10^{-1}$, $R = 1$ mm, $\varepsilon = 60$ keV, $k_C \sim 9 \cdot 10^9$ Nm$^2$/C$^2$ gives a value $N = 2.6 \cdot 10^6$. In [25] was demonstrated that diffraction pattern of sulfur hexafluoride molecules can be obtained with picosecond electron pulses containing $N \sim 10^6$ electrons per pulse and the uncertainty for the determined the internuclear distances of $\sim 0.1\%$.

We also obtain the expression for $N_{max}$ as a function of the electron pulse duration, defined as $\tau_e = 2l / v_z$, where $v_z$ – velocity of the electron bunch. Using (7), (8) and (11), we find:

$$N_{max} = 2\sqrt{2\pi} \frac{R\kappa\varepsilon}{k_C e^2} \tilde{\theta}^2 \begin{cases} \dfrac{\sqrt{\xi^2 - 1}}{\operatorname{arcosh} \xi}, & \text{for } \xi > 1 \\ \dfrac{\sqrt{1 - \xi^2}}{\arccos \xi}, & \text{for } \xi < 1 \end{cases}, \quad (24)$$

where $\xi = v_z \tau_e / 2R$. This dependence is shown in Figure 4.

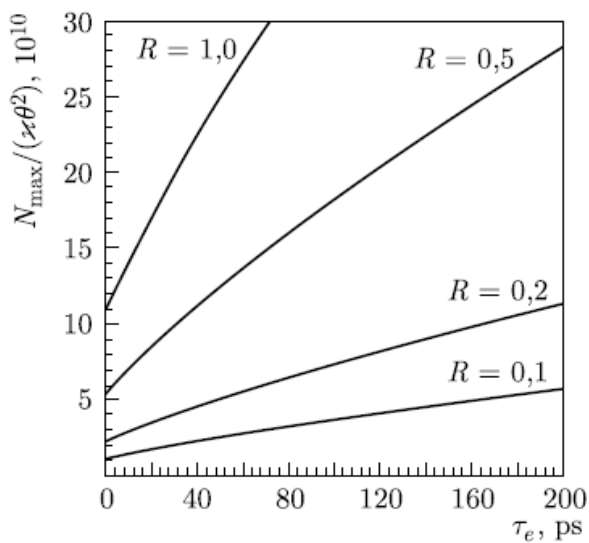

**Fig. 4.** The dependence of the maximum number of electrons in the pulse on its duration for a given error in the determination of the structural parameters $\delta r / r = \kappa = 0.001$ for four values of the transverse dimension of R (mm).

The obtained general relations allow of estimating of systematic errors in determining the dynamic molecular parameters caused by the Coulomb interactions in the electron bunch as a function of the pulse duration and the number of particles in the electron bunch.

## 3. Conclusions

The electron diffraction with a high time resolution has opened the possibility of direct observation of the processes occurring in the transient state of the substance. Here it is necessary to provide a temporary resolution of the order of 100 fs, which corresponds to the transition of the system through the energy barrier of the potential surface, which describes the chemical reaction - the process of the breaking and the formation of new bonds between the interacting agents. Thus it can be opened the possibility of the investigation of the dynamics of coherent nuclear molecular systems and the condensed matter [8-11]. In the past two decades it has been possible to observe the nuclear motion in the temporal interval corresponding to the period of the nuclear oscillation. The observed coherent changes in the nuclear system at such temporal intervals determine the fundamental shift from the standard kinetics to the dynamics of the phase trajectory of a single molecule, the molecular quantum state tomography [9-11, 28].

At present, the method of ultrafast diffraction is intensively developing. The great opportunities for the study of the structural dynamics are opened by 4D methods of ultrafast electron crystallography and electron microscopy with high temporal resolution from micro- to femtoseconds (see review articles [5-7, 20,22,29,30]. The recent advances in the formation of ultrashort electron pulses allow us to go to the area of attosecond temporal resolution and to observe the coherent dynamics of the electrons in the molecule [31-33].

It is rather impressive, that laser alignment of the molecules in the gas phase helps to restore the molecular structure, by observing the multiple patterns of the ultrashort electron beam diffraction [34]. In these conditions, in principle, no additional parameters are necessary for the elucidating of complex molecular structures and intermediate states in molecular dynamics. It should be noted, that it [34] was the first TRED experiment with subpicosecond temporal resolution done in the gas phase. In this experiment the molecules $CF_3I$ were aligned impulsively with a femtosecond laser pulse and probed with a 500-fs electron pulse two picoseconds later, when the degree of the molecular alignment reaches a maximum.

## 4. Acknowledgements

This work was supported by the Russian Foundation for Basic Research (RFBR) Grant No. 13-02-12407 OFI_m2 and No. 12-02-00840-a.